\documentclass[12pt]{article}
\usepackage{latexsym}
\usepackage{geometry} 
\usepackage{pict2e}

\def\bs{\begin{subequations}}
\def\es{\end{subequations}}

\catcode`\@=11

\newtoks\@stequation
\def\subequations{\refstepcounter{equation}
  \edef\@savedequation{\the\c@equation}%
  \@stequation=\expandafter{\theequation}
  \edef\@savedtheequation{\the\@stequation}
  \edef\oldtheequation{\theequation}%
  \setcounter{equation}{0}%
  \def\theequation{\oldtheequation\alph{equation}}}

\def\endsubequations{\setcounter{equation}{\@savedequation}%
  \@stequation=\expandafter{\@savedtheequation}%
  \edef\theequation{\the\@stequation}\global\@ignoretrue}

\usepackage{hyperref}

\begin{document}
\begin{titlepage}
\begin{center}
 {\bf {Tachyon Interactions} }
 
 Charles Schwartz, Department of Physics \\University of California, Berkeley, CA. 94720\\
 schwartz@physics.berkeley.edu   \\
 
 January 8, 2023
  
  \end{center}
\vskip 1cm
Accepted for publication by the MDPI journal, Symmetry, Special Issue,

"Symmetry in Quantum Fields, Gravitation, and Cosmology"
 
 \vskip 1cm
 
\vfill

\begin{abstract}
A consistent theory of free tachyons has shown how tachyon neutrinos can explain major cosmological phenomena, Dark Energy and Dark Matter.  Now we investigate how tachyon neutrinos might interact with other particles: the weak interactions. Using the quantized field operators for electrons and tachyon neutrinos, the simplest interaction shows how the chirality selection rule, put in by force in the Standard Model, comes out naturally. Then I wander into a re-study of what we do with negative frequencies of plane wave solutions of relativistic wave equations. The findings are simple and surprising, leading to a novel understanding of how to construct quantum field theories.

\end{abstract}

\vskip 2cm
\begin{center}
{\textbf{Advisory}}

Any editor or reader who harbors old prejudice against the idea of tachyons is invited to read these notes:  
\url{https://www.ocf.berkeley.edu/~schwrtz/debunk2.pdf} 
\end{center}

\vfill

\end{titlepage}

\section{Introduction}

Having developed a consistent theory of free tachyons ("faster-than-light" particles) and shown how tachyon neutrinos can explain major cosmological phenomena (Dark Energy and Dark Matter),\cite{CS1} I now want to investigate how tachyon neutrinos might interact with other particles.  The weak interactions come immediately to mind.

The Standard Model asserts that neutrinos interact with other leptons through a purely chiral coupling, at least at high energies. Let's see what a tachyon spin 1/2 theory for neutrinos would imply.  Section 2 is a review of the quantized field theory for a Dirac particle, either a tachyon or an ordinary  (slower-than-light) particle. Section 3 introduces a simple interaction between the two types of particles and we see the result of only two assumptions -  Lorentz invariance and Lepton Number conservation: the chirality selection rule comes out automatically. In Section 4 I ask, "Is this really new?"

In Section 5 I re-examine the old question of how to work with negative frequency solutions of relativistic wave equations. Going back to the even older conundrum of wave-particle duality, I take a close look at building wave packets for relativistic particles. This leads to a surprisingly simple rule for matching the parameters $k^\mu$ of a plane wave to the parameters $p^\mu$ of a relativistic particle. This approach appears to be something fundamentally correct for relativistic quantum theory - and strangely missing from conventional pedagogy of that theory. In Section 6 we see that this gives us a new symmetry between ordinary particle Dirac theory and tachyon Dirac theory.

\section{ Quantized fields}

Let's start with the density of states on the mass shell, for plane wave solutions of any relativistic wave equation.
\begin{eqnarray}
&&\int d^4k\; \delta(k^\mu k_\mu - \epsilon m^2) = \\
ordinary\; \epsilon = +1: &&\sum_{s = \pm 1} \int \;d^3k/2\omega, \;\;\; \omega = +\sqrt{k^2+m^2}, \;\; k_0 = s \;\omega, \\
tachyon\; \epsilon = -1:\;\;\;&& \int_{-\infty}^\infty d\omega\; \int d^2\hat{k}/2k, \;\;\; k = +\sqrt{\omega^2 + m^2}.
\end{eqnarray}
Here we have separated the momentum 3-vector $\textbf{k} = k\; \hat{k}$ into a magnitude and a direction (for tachyons); and the parameter $s = \pm 1$ designates the two sectors of the mass shell (for ordinary particles)
 
We are interested in spin 1/2 relativistic particles obeying two versions of the Dirac equation. The wavefunctions separate into a plane wave in the spacetime coordinates, $x^\mu$, with the 4-vector of parameters $k^\mu = (\omega, \textbf{k})$ and a 4-component algebraic spinor $u \; or\; v$.
\begin{eqnarray}
ordinary\; : \;\;\;&& i\gamma^\mu \partial_\mu \psi = m \psi, \;\;\; \psi = e^{-ik_\mu x^\mu}\;u(\textbf{k}), \\
&&u_{s,h}(\textbf{k}) = 1/\sqrt{2|s\;\omega+m|} \left(\begin{array}{c} s\; \omega+m \\ h\;k \end{array} \right)\; |\hat{k}, h>, \\ \nonumber \\
tachyon\;: \;\;\;&&i\gamma^\mu \partial_\mu \psi = i\;m \psi, \;\;\; \psi = e^{-ik_\mu x^\mu}\;v(\omega, \hat{k}), \\ 
&&v_h(\omega, \hat{k}) = 1/\sqrt{2k}\;\left( \begin{array}{c} \omega + im \\ h\; k \end{array}\right)\;|\hat{k}, h>.
\end{eqnarray}

Here $|\hat{k}, h>$ is a spin  eigenfunction with helicity $h = \pm 1$ relative to the direction of the momentum $\hat{k}$.

Now we are ready to assemble these parts, together with annihilation and creation operators, into the quantized field operators. For the ordinary Dirac particle, call it an electron, it is the number $s = \pm 1$
which we originally introduced to designate each half of the bifurcated mass shell, that now serves to choose which function goes with which operator.

\begin{equation}
\psi_e (x) = \int  d^3k\; e^{i\textbf{k}\cdot\textbf{x}}\;\sum_{h=\pm 1} [e^{-i \omega t}a_{s=+1, h}(\textbf{k})\; u_{s=1, h}(\textbf{k})  + e^{+i \omega t} a^\dagger_{s=-1, h}(\textbf{k})\;u_{s=-1, h}(\textbf{k}) ].\label{b8}
\end{equation}
The operators a annihilate the vacuum, $a_{s, h}(\textbf{k})\;|0> = 0$; and they are labelled as are the functions u. Those operators obey an anti-commutator algebra:

\begin{equation}
[a_{s,h}(\textbf{k}), a^\dagger_{s', h'}(\textbf{k}')]_+ =\delta^3(\textbf{k} - \textbf{k}')\;\delta_{s, s'}\;\delta_{h, h'}.\label{b9}
\end{equation}

 If we calculate the anticommutator of the electron fields at two spacetime points we get the result that it vanishes for spacelike separation of those points.That is the familiar statement, called "causality", for slower-than-light particles.
 
The quantized field for a Dirac tachyon, let us call it a neutrino,  looks like this. \cite{CS3}
\begin{eqnarray}
\psi_\nu (x) = \;\;\;\;\;\;\;\;\;\; \nonumber\\
\int _{-\infty}^{+\infty} d \omega \;e^{-i\omega t}\int d^2 \hat{k}\; k^2 e^{i \textbf{k}\cdot\textbf{x}} [b_{h=-1}(\omega, \hat{k})\; v_{h=-1}(\omega, \hat{k}) \; + \;b^\dagger_{h=+1}(\omega, \hat{k}) \; v_{h=+1}(\omega, \hat{k}) ].\label{b10}
\end{eqnarray}
Here the b's are annihilation operators,  $b |0> = 0$, and the $b^\dagger$'s are creation operators;  the v's are Dirac 4-component spinors given above. The anti-commutator algebra for the tachyons is,
\begin{equation}
[b_h(\omega, \hat{k}), b^\dagger_{h'}(\omega', \hat{k}')]_+ = \delta(\omega - \omega') \frac{\delta^2(\hat{k} - \hat{k}')}{k^2}\;\delta_{h, h'}.\label{b11}
\end{equation}

Using the above formulas we calculate the anticommutator for the fields at two spacetime points:
\begin{equation}
[\psi_\nu^\dagger(x), \psi_\nu(x')]_+ = {\cal{D}}\; \int_{-\infty}^{+\infty} d\omega \;e^{-i\omega (t-t')} k^2 \frac{sin kr }{kr}=0 \;for \;|t-t'| >r, \;\;\; r=|\textbf{x} - \textbf{x}'|,
\end{equation}
where ${\cal{D}}$ is a set of differential operators acting on the spacetime coordinates. 
This result confirms that we have a correct theory for tachyons: tachyons are defined by the distinction that the particles (fields) always travel (propagate) outside of the lightcone.

Note that in the first case the plane waves have opposite phase for the two types of operators while in the second case they are the same. Is this correct? To explore this question let me construct a representation of the displacement operators $P^\mu$ in each case.  First, for ordinary particles,
\begin{equation}
P^\mu = \int d^3 k\; \sum_{s,h}\; k^\mu \;a^\dagger_{s, h} (\textbf{k})\; a_{s, h} (\textbf{k}), \;\;\;\;\;
\;[P^\mu, \psi_e(x)]_{-}= i \partial^\mu \psi_e (x).\label{b13}
\end{equation}
The second equation is readily verified using the definitions above. Now, for the tachyon field, as constructed above, I need to construct it a bit differently.
\begin{equation}
P^\mu = \int d \omega\;\int d^2\hat{k}\;k^2\; \sum_h\;(-h)\; k^\mu\; b^\dagger_h (\omega, \hat{k})\; b_h(\omega, \hat{k}).
\end{equation}
That extra factor h makes it perform as it should when this acts upon the field operator $\psi_\nu$; but this gives an unexpected additional detail for distinguishing the two types of Dirac particles. (Could this factor h be related to the indefinite metric that I introduced in the Wigner analysis of the Little Group for spin 1/2 tachyons?)

One should continue to construct all the other operators of the Poincare Group in this fashion, looking to see if this is all consistent (and unique?) I leave this work to later.

The reader is encouraged to look over the above sets of formulas and note where the two types of Dirac spinor fields are similar and where they are sharply different.  The most striking difference is noting which quantum numbers are used to identify the particle vs the antiparticle (which quantum numbers go with the annihilation operators and which with the creation operators). This pivotal role is played by the number s for electrons (designating which sheet of the mass shell we are looking at) and it is played by the helicity h for the tachyon. I assert that, in each case, this is a Lorentz invariant rule. 

What do we know about Lorentz transformations and the Dirac equation? The spacetime coordinates (or their derivatives) are changed in the classical way; and the four component Dirac spinor is multiplied by an algebraic matrix. It is exactly the same process for any value of the mass in the Dirac equation, including the change from a real mass to an imaginary mass (for the tachyon case). We are talking, for now, about proper, orthochronous Lorentz transformations. Questions about the discrete operations - parity, time reversal and charge conjugation - may come later.

We know very well that the choice of one or the other sector of the mass shell for ordinary particles is Lorentz invariant. How do we know that the helicity is a Lorentz invariant label for tachyons? Here is a simple table of calculations done using the Dirac spinors for simple plane wave solutions of the two types of Dirac equations. The first row is known as a scalar and the second row is called a pseudoscalar; they are all invariant under p-o Lorentz transformations. [$\bar{\psi} = \psi^\dagger \gamma^0$]
\begin{eqnarray}
(\bar{\psi}_e , \psi_e) = s\; m\;\;\;\;\;\;\;\;\;\; (\bar{\psi}_\nu, \psi_\nu) = 0 \\
i(\bar{\psi}_e , \gamma_5\;\psi_e) = 0. \;\;\;\;\;\;\;\;\;\; i(\bar{\psi}_\nu , \gamma_5\;\psi_\nu) = -h\; m.
\end{eqnarray}
We see here that the helicity is indeed a Lorentz invariant number; and we also see a sort of symmetry between the roles of s and h. Further comparisons for higher rank tensors are given in reference \cite{CS3}

This identification leads directly to the definition of Lepton Number for each type.

\begin{equation}
N_{tachyon} = \int _{-\infty}^{+\infty} d \omega \int d^2 \hat{k}\; k^2 \;\Sigma_h\; (-h)\;b^\dagger_{h}(\omega, \hat{k}) \;b_{h}(\omega, \hat{k}) .
\end{equation}
\begin{equation}
N_{ordinary} = \int d^3 k \;\Sigma_h\;\Sigma_s\; s\; a^\dagger_{s, h}(\textbf{k})\;a_{s, h} (\textbf{k}).
\end{equation}
Here $[N, \psi]_{-} = -\psi$ for either field.

\section{Interaction}

Now we are ready to construct a possible interaction between the tachyon neutrino and the ordinary electron. A 4-vector coupling is very popular:
\begin{equation}
{\mathcal{L}} = \psi^\dagger_e(x) \gamma^0\;\gamma^\mu\; \psi_\nu (x) \; B_\mu (x)+ H.C.\label{c1}
\end{equation}
And $B_\mu$ is some vector field. This is covariant under proper orthochronous Lorentz transformations. This pairing of the two types of fields will preserve the total Lepton Number $N_t + N_o$. The above interaction will not allow for neutrinoless double beta decay; that agrees with the latest experiments on this topic.\cite{CU}

What we know from experimental physics and have built into the Standard Model is that the weak interactions conserve chirality. That means, in the usual formalism, that the projection operator $(1\pm \gamma_5)$ sits inside the mixed current. The tachyon states are labeled by helicity, which is not the same as chirality; but, in fact, those two labels become the same as the particle states approach the light cone. ALL (really) experiments with neutrinos are at energies in the range of MeV or much higher. With the tachyon neutrinos taken to have a very small mass, around 1 eV or less, this means that a tachyon neutrino wave function is very very close to a chiral egen-state.The coupling shown above \emph{automatically} gives us the result that electrons engage in the weak interactions through a chiral projection operator. In the Standard Model this is put in by force; in the tachyon theory this comes naturally.

This appears to be a new and significant discovery in theoretical physics. Future work would ask about fitting this into the full Standard Model and dealing with mass mixing of neutrinos.

\section{ But is this really new?}

Several other authors have written about neutrinos being tachyons, \cite{CHK} \cite{JW} \cite{RCC}. They all find a connection between the tachyon spin 1/2 wave functions and the chirality of weak interactions. However their theories of how to quantize a Dirac tachyon wave equation follow too closely the path used for ordinary particles. 
I quote from the abstract of \cite{JW},

"The quantum field theory of superluminal (tachyonic) particles is plagued by a number of problems, which include the Lorentz non-invariance of the vacuum state, the ambiguous separation of the field operator into creation and annihilation operators under Lorentz transformations, and the necessity of a complex reinterpretation principle for quantum processes. Another unsolved question ..."

The theory presented in this paper does not have such problems.

\section{Dealing with negative frequencies}

There are serious questions with the above formulation that I must now face up to. One question is, How do you interpret negative frequency wavefunctions?  I have previously given a neat answer to that question for a classical tachyon: reverse the sign of $k^\mu$ and interchange the categories of "in-states" and "out-states". (See Reference \cite{CS1}, Section 2.3) I need now find the correct path to do this in quantum mechanics.

This turns out to be a tussle with words more than a problem in mathematics or in physics.  It brings us back to a very ancient conundrum in the history of the quantum theory:  What is the relation between a wave and a particle? We start with a wave equation (Klein Gordon or Dirac or ...) and we select solutions that are eigenfunctions of the spacetime displacement operators (the generators $P^\mu$ of the Poincar\'e group).
\begin{equation}
P^\mu \;e^{ik_\mu x^\mu} = k^\mu \;e^{ik_\mu x^\mu}.
\end{equation}
But a plane wave is in no sense localizable. A particle, on the other hand, is something that is, at least somewhat, localizable in spacetime; and we represent it classically as a line trajectory in 4-dimensions. For quantum theory we do not ask for point-like localizability; but we must have something that makes  sense when we call it a particle.  The answer lies in the exercise of building a wave packet.

\begin{equation}
\psi(x) = e^{\pm i(\mathbf{k} \cdot \textbf{x} - \omega t)} \rightarrow\; e^{\pm i(\mathbf{k} \cdot \mathbf{x} - \omega t)}\;F(\mathbf{x}, t)
\end{equation}
The function F() is the envelope of the wave packet; it varies very slowly in space-time compared to the rapid oscillations of the plane wave factor. We now require that this construction be a solution of the Klein-Gordon equation. The operator we apply to this wave packet is $\partial^2_t - \sum_i \partial^2_{x_i} \pm m^2$. Acting only on the plane wave factor this gives zero; the interesting term is when the second derivatives act once on the plane wave and once on the envelope.  This gives us the determining equation,
\begin{equation}
k_0\;\partial_t \; F() - \sum_i\; k_i\; \partial_{x_i}\; F()= 0.
\end{equation}
and the solution is, 
\begin{equation}
F() = F(\textbf{x} - \textbf{V}\;t), \;\;\;\;\; \textbf{V} = \textbf{k}\; /\omega, \;\;\;\;\; k^\mu = (\omega, \textbf{k}).
\end{equation}
This, at first, looks like a very familiar story about the group velocity. The physical particle, described by the wave packet envelope, moves through space-time with a velocity $\textbf{V}$. We have the familiar way of describing the trajectory of a classical particle in Special Relativity according to a four-vector $p^\mu = (E, \textbf{p})$; and we now see the simple correspondence betweeen wave and particle - an equation we physicists teach to students in introductory courses on the quantum theory:
\begin{equation}
p^\mu\; (for\;the\; particle) = k^\mu \;(for\;the\;wave.).
\end{equation}
(We may put a factor $\hbar$ in this equation; and then, in our sophistication choose units such that $\hbar = 1$.) This is not just neat, it fits the equation above about the group velocity. The familiar representation for a free particle is, $p^\mu = (m\gamma, m \gamma \textbf{v})$, where $\gamma = 1/\sqrt{|1-v^2/c^2|}$; and this says $v = p/E$.

BUT WAIT!

The equation we got from studying wave packets allows for negative frequencies $\omega < 0$; and in that situation the corresponding free particle is moving in the direction \emph{opposite} the direction of the wave vector $\textbf{k}$. [Note. In this section the use of the word "particle" is not meant to distinguish particle from antiparticle.]

For ordinary particles, as we have written above, $\omega$ is always positive. But for tachyons, we are now looking at the states of negative $\omega$; and this is our conclusion:

RULE 1. For Tachyons, a negative frequency plane wave corresponds to a particle which travels in the direction opposite to the wave propagation vector $\textbf{k}$.

Since we are looking closely at minus signs, what about the sign of the energy? If $\omega$ can be negative and we like the association of the two fourvectors $k^\mu$ and $p^\mu$, then do we face negative energies?  I am now going to recite another rule.

RULE 2.  For any free particle, the energy is defined to be a positive quantity: $E = p^0 > 0$. 

I do not offer any theoretical derivation of this rule nor do I suggest any experiments to verify this rule.  I offer it as a common habit, an assumption, a linguistic definition - perhaps an axiom - of particle physics.  I do not claim this is an original idea of mine; I have seen such statements in the writings of others, typically in order to prove some mathematical theorem.  In fact, in my earlier studies of tachyons I have consciously felt that such a rule must be wrong.  But now I change my opinion, in the context of this discussion where we are making clear separation between the language of wavefunctions and the language of particles.

The two rules above can be nicely combined,

RULE 3.  For Tachyons, a negative frequency plane wave corresponds to a particle which has the four-momentum $p^\mu = - k^\mu$.

This conclusion may be said to be trivial, looking at the simple (and familiar) calculation we used to arrive at it. Yet I take it as a rather profound statement about the quantum theory of tachyons. Other tachyon authors, mentioned in Section 4 of this paper, have projected out all negative frequency solutions of the wave equation; thus they avoid this (strange) situation but the cost is the destruction of Lorentz invariance.  My approach, has scrupulously preserved Lorentz transformations, the whole symmetry of the Poincar\'e group; and we now will learn to live with this strange new arrangement.

One lesson is to be clear that there are different meanings to the parameters $k^\mu$ used to identfiy free particle solutions of a wave equation and the parameters $p^\mu$ used to identify kinematics of a free particle. For ordinary particles these two are always identical; but for tachyons, these two may differ by a minus sign. This (strange = unfamiliar) behavior depends entirely upon the sign of the frequency $\omega$. Yet, I claim, we have full Lorentz invariance for the theory. The process of making a correspondence between particle and wave attributes may well depend on what Lorentz frame we sit in when we do this. However, what the Special Theory of Relativity says is that many details of any physical experiment/observation may vary from one reference frame to another but the fundamental Laws of Physics should be covariant in their mathematical form.  Such Laws are, for example, the overall conservation of total energy and momentum for any isolated system.

One of the issues that needs attention here is the multiplicity of ways in which the word "energy" has been employed in physics. So long as all particles were assumed to travel within the lightcone, those many meanings all converged. Tachyons make this a more demanding issue. I should mention the Energy-Momentum tensor, which, in my earlier writings about tachyons, may have plus or minus signs in front that nobody had ever seen before.

\subsection{Another new Rule}
There is one more issue we must attend to in figuring out how to map the mathematics of wave functions onto the mathematics of particles. In my earlier analysis (see section 2.3 in reference \cite{CS1}) of classical particles and learning how to handle negative values of the energy $E = p^0$ in the 4-vector $p^\mu$ it was found that not only did one have to change $\textbf{p} \rightarrow \;-\textbf{p}$ (along with making $E > 0$), but one also had to interchange the designations of "in-states" and "out-states". That was done in the context of a general scattering process; but here we are simply trying to construct the quantum description of free relativistic particles. How do we carry over that idea to our present work?

The answer lies in the characterization of "creation" operators and "annihilation" operators.  We have before us what looks at first like two different sets of words: "in-states" vs "out-states" on the one hand; and "creation" vs. "annihilation" on the other hand. [Please pardon the mixed metaphors: "words" on "hands".] That difference is resolved when we note that each pair of words depends on our visualization of physical processes evolving as time moves forward. Once again, I would evoke the notion of an "axiom of physics" to justify this identification.

RULE 4. When constructing the quantized field operator for any type of particle - ordinary or tachyon - the coefficients correlated with $k^0 > 0$ wavefunctions are annihilation operators and those correlated with $k^0 < 0$ wavefunctions are creation operators.

This new Rule is nothing new for ordinary particles; but it is game-changing for tachyons. It means that I must break the set of tachyon wavefunctions, originally written as a continuous whole $-\infty < k^0 < +\infty$, into two separate sets: $k^0 = s \;\omega$, where $\omega > 0$ and $s = \pm 1$. This seems to put me dangerously close to the work of those other authors, cited in Section 4, which I disparaged. So, here is the saving grace. 

\begin{equation}
\int_{-\infty} ^{+\infty}\; d\omega\; f(\omega) = \sum_{s= \pm 1}\;\int_0^{+\infty}\; s\; d \omega \;f(s\omega). \label{e9}
\end{equation}
The left hand side of this equation was used to cover the whole of the tachyon mass shell; and this way of writing made the formalism manifestly Lorentz invariant. The right hand side seems to threaten Lorentz invariance; yet the two forms are mathematically identical. Remember that showing Lorentz invariance is a mathematical task carried out upon the spacetime coordinates and the accompanying Dirac spinors; the mapping of wavefunctions to particles is a separate (physical) task that has some human prejudice built into it: regarding how we interpret $s=-1$ states. There is a discontinuity in this mapping, which we saw first in the analysis of relativistic wave packets; and this does not break our fundamental symmetry if Lorentz invariance.

\subsection{Application of the new Rules}

Let's look at the formulas of Section 2. First, consider Equation (\ref{b8}) for the field operator of an ordinary Dirac particle. The first terms, involve the annihilation operators $a_{...}$ along with the positive frequencies $e^{-i\omega t}$, while the second terms involve the creation operators $a^\dagger_{...}$ along with the negative frequencies $e^{i\omega t}$. This has the requirement of Rule 4 already built in. Now we construct one-particle states as they are created from the vacuum state $|0>$.
\begin{equation}
a^\dagger_{s=+1, h} (\textbf{k}) \;|0> = |A>, \;\;\; a^\dagger_{s=-1, h} (\textbf{k}) |0> = |B>;
\end{equation}
and calculate some inner products in the Fock space, which represents these particles.
\begin{eqnarray}
<0|\;\psi_e(x) |A> = e^{i\textbf{k} \cdot \textbf{x}}\;e^{-i\omega t}\;u_{s=+1, h} (\textbf{k}), \\
<A|\;\psi_e(x)\;|0> = 0, \\
<B|\;\psi_e(x)\;|0> = e^{i\textbf{k} \cdot \textbf{x}}\;e^{+i \omega t}\;u_{s=-1, h} (\textbf{k}), \\
<0|\;\psi_e(x)\;|B> = 0.
\end{eqnarray}
Remember that here $\omega > 0$ and all particle states are defined to have positive energy. The first equation above tells us: the state $|A>$ is what we call the particle and it has momentum $\textbf{p} = \textbf{k}$, with helicity h; this also says that the quantized field operator $\psi$ annihilates the particle.
The second equation above tells us that $\psi$ cannot create this one particle state A. The third equation above shows that the state $|B>$ derives from the wavefunction with negative frequency and momentum $\textbf{k}$. Therefore, following Rule 3, we identify the state B as an "antiparticle" with momentum $\textbf{p} = - \textbf{k}$. This same equation also tells us that the field operator $\psi$ creates such an antiparticle. The fourth equation above says that $\psi$ cannot annihilate this antiparticle.  I expect that this is all familiar talk for ordinary particles; what we have added to the pedagogy is a firm understanding, from our analysis of wave packets, of the logic of this formalism.

Now we turn to tachyons and Equation (\ref{b10}). I will rewrite that equation, making use of (\ref{e9}).

\begin{eqnarray}
\psi_\nu (x) = \int_0^{+\infty}\; d\omega\;\int d^2 \hat{k}\; k^2\; e^{i \textbf{k}\cdot\textbf{x}} \;\;\times \;\;\;\;\;\nonumber\\
\;[e^{-i\omega t}\;b_{h=-1}(\omega, \hat{k})\; v_{h=-1}(\omega, \hat{k}) \; - e^{+i \omega t}\; b^\dagger_{h=-1}(-\omega, \hat{k})\;v_{h=-1}(-\omega, \hat{k}) + \nonumber\\
e^{-i\omega t}\;b^\dagger_{h=+1}(\omega, \hat{k})\; v_{h=+1}(\omega, \hat{k}) \; - e^{+i \omega t}\; b_{h=+1}(-\omega, \hat{k})\;v_{h=+1}(-\omega, \hat{k}) ]
\end{eqnarray}
This differs from (\ref{b10}) in the assignment of $b_{...}$ and $b^\dagger_{...}$ operators: I follow Rule 4 and also follow the naming of the helicity $h = \pm 1$ as signifying particle vs antiparticle. Also remember that $\omega$, in this new formulation, is a strictly positive variable.

Now construct the one-particle tachyon states:
\begin{equation}
b^\dagger_{h}(\omega, \hat{k}) |0> = |A_h>, \;\;\;\;\;b^\dagger_h(-\omega, \hat{k})|0> = |B_h>, \;\;\; h = \pm 1,
\end{equation}
and calculate the set of inner products involving the field operator.
\begin{eqnarray}
<0|\; \psi_\nu(x)\;|A_{h=-1}> =  e^{i\textbf{k}\cdot\textbf{x} - i \omega t} v_{h=-1}(\omega, \hat{k}), \;\;\;      <A_{h=-1}|\psi_\nu(x)|0> = 0, \\
<0|\; \psi_\nu(x)\;|A_{h=+1}> = 0, \;\;\; <A_{h=+1}|\psi_\nu(x)|0> = e^{i\textbf{k}\cdot\textbf{x} - i \omega t} v_{h=+1}(\omega, \hat{k}), \\
<0|\; \psi_\nu(x)\;|B_{h=-1}> = 0, \;\;\;   <B_{h=-1}|\psi_\nu(x)|0> = - e^{i\textbf{k}\cdot\textbf{x} + i \omega t} v_{h=-1}(-\omega, \hat{k}), \\
<0|\; \psi_\nu(x)\;|B_{h=+1}> = - e^{i\textbf{k} \cdot \textbf{x}+ i \omega t} v_{h=+1}(-\omega, \hat{k}), \;\;\;       <B_{h=+1}|\psi_\nu(x)|0> = 0.
\end{eqnarray}
We read these equations as follows. State $A_{h=-1}$ is a "particle" with momentum $\textbf{p} = \textbf{k}$ and it is annihilated by $\psi_\nu$. State $A_{h=+1}$ is an "antiparticle" with momentum $\textbf{p} = \textbf{k}$ and it is created by $\psi_\nu$. State $B_{h=-1}$ is a "particle" with momentum $\textbf{p} = - \textbf{k}$ and it is created by $\psi_\nu$. State $B_{h=+1}$ is an "antiparticle with momentum $\textbf{p} = - \textbf{k}$ and it is annihilated by $\psi_\nu$.

Note, from the above bookkeeping, that there are two distinct states for each "particle" and for each "antiparticle" of any given momentum $\textbf{p}$. For ordinary particles those two states are distinguished by the helicity $h = \pm 1$. For tachyons those two states are distinguished by the sign of the frequency $s = \pm 1$ in the corresponding wavefunction.

\section{Summary}
I want to display the counting of states for a free Dirac wavefunction/particle. The wavefunctions are characterized by four parameters of the plane wave assembled in the vector $k^\mu = (\omega, \textbf{k})$ and also the spin, aligned to the momentum as helicity $h = \pm 1$. There are two primary cases: ordinary particles (moving always slower than light); and tachyons (always traveling faster than light).

For $k^\mu k_\mu = + m^2$.
There are 2 distinct entities, called "particle" and "anti-particle", indicated by $s=\pm 1$, the sign of the frequency $k^0$.
Each of these has 2 quantum states indicated by the helicity h.

For $k^\mu k_\mu = -m^2$.
There are 2 distinct entities, called "particle" and "anti-particle", indicated by the helicity $h=\pm 1$. 
Each of these has 2 quantum states indicated by s, the sign of the frequency $k^0$.

In both cases we see that Lorentz transformations cannot change the first category but can change the second category.

This is a very neat picture of the wave-particle correspondence for the Dirac equation. What is new here? The negative frequency solutions for ordinary particles have long been interpreted as describing the anti-particle; but the reasons given are worth looking at. (Historically it has a lot to do with electric charge, the Pauli principle, and, ultimately, the actual discovery of the positron.) For tachyons, it has been customary (in my own prior work) to consider the whole mass shell $-\infty < \omega < + \infty$ as an indivisible class. Others authors have separated out negative frequencies and eliminated them from the physical states (thus violating Lorentz invariance). The view presented here is something new. I find this quite satisfying, aesthetically: it displays the full Symmetry of exchanging the roles of the two parameters $s= \pm 1$ and $h=\pm 1$ as one switches between ordinary particles and tachyons. 

Is this really the discovery of a new internal quantum number for neutrinos? Does it deserve a name? How might it be detected physically?

\section{Conclusions}

The material presented in Sections 5 and summarized in Section 6 appears to be something so fundamental to quantum field theory that  physicists ought to have worked this out many years ago. Yet, I am not aware of any other authors who have published such analysis of wave packets from relativistic wave equations. Perhaps the Reviewers and Editors who pass judgment on this paper will enlighten me on this question.

The main takeaway is that the 4-vector $k^\mu$, which characterizes plane wave solutions of a relativistic wave equation, should be seen as distinct from the 4-vector $p^\mu$, which characterizes quanta of the field - those things that we call free particles (or anti-particles). The mapping of one 4-vector onto the other is not trivial; but it turns out to be simple and sensible; and we also learn how Lorentz invariance is preserved.

For the case of ordinary Dirac particles, this new insight may be seen as no more than an improvement in pedagogy. However, for the case of Dirac tachyons, the implications are substantial; and the resulting reformulation is gratifying.

The interaction of tachyon neutrinos with electrons, as presented in Section 3 should be the basis for future study: calculate matrix elements for specific beta decay processes and see if there are suggestions for detailed experiments that could verify, or negate, this theory.

\end{document}